\documentclass[aps,prb,twocolumn,superscriptaddress,floatfix,showpacs]{revtex4}
\usepackage{epsfig,amsmath,amssymb,color,graphicx,units}
\newcommand{\jjj}{$J_1$-$J_2$-$J_3$ }

\begin{document}

\title{Quantum Phases of the Planar Antiferromagnetic
  $\mathbf{J_1}$-$\mathbf{J_2}$-$\mathbf{J_3}$ Heisenberg-Model}

\author{Johannes Reuther}
\email[E-mail: ]{reuther@tkm.uni-karlsruhe.de}
\author{Peter W\"olfle}
\affiliation{Institut f\"ur Theorie der Kondensierten Materie, Karlsruhe
Institute of Technology, 76128 Karlsruhe, Germany}

\author{Rachid Darradi}
\author{Wolfram Brenig}
\affiliation{Institut f\"ur Theoretische Physik, Technische
Universit\"at Braunschweig, 38106 Braunschweig, Germany}

\author{Marcelo Arlego}
\affiliation{Departamento de F\'isica, Universidad Nacional de La Plata,
C.C. 67, 1900 La Plata, Argentina.}

\author{Johannes Richter}
\affiliation{Institut f\"ur Theoretische Physik, Universit\"at Magdeburg, 39016
Magdeburg, Germany}

\date{\today}

\begin{abstract} We present results of a complementary analysis of the
frustrated planar $J_1$-$J_2$-$J_3$ spin-1/2 quantum-antiferromagnet.  Using
dynamical functional renormalization group, high-order coupled cluster
calculations, and series expansion based on the flow equation method, we have
calculated generalized momentum resolved susceptibilities, the ground state
energy, the magnetic order parameter, and the elementary excitation gap.  From
these we determine a quantum phase diagram which shows a large window of a
quantum paramagnetic phase situated between the N\'eel, spiral and collinear
states, which are present already in the classical $J_1$-$J_2$-$J_3$
antiferromagnet. Our findings are consistent with substantial plaquette
correlations in the quantum paramagnetic phase. The extent of the quantum
paramagnetic region is found to be in satisfying agreement between the three
different approaches we have employed.  \end{abstract}


\maketitle

\section{Introduction}

The search for exotic quantum phases is one of the main interests in the study
of spin systems with competing interaction. Ultimately this search may uncover
spin liquids (SL) without any magnetic order or long range correlations.
En-route however, many interesting quantum paramagnets (QP) lie, which are
not magnetically ordered, however exhibit broken spatial symmetries with
respect to short range magnetic correlations, either spontaneously or by virtue
of the lattice structure, i.e. valence bond crystals (VBC) or solids (VBS). In
two dimensions a paradigmatic system in this context is the antiferromagnetic
(AFM) $J_1$-$J_2$ model on the square lattice with frustrating diagonal
exchange.  As a function of the single parameter $j=J_2/J_1$ this model is
widely accepted to undergo a transition from a N\'eel state at $j\lesssim 0.4$,
to a QP phase for $0.4 \lesssim j \lesssim 0.6$ and to a collinear AFM phase
beyond that. However, even two decades after first analysis of 
this \cite{Chandra1988,Schulz1992}, no consensus has been reached on the nature
of the QP phase and the type of transition into it, see
e.g. Ref.~\onlinecite{Richter2010} and references therein. Possible QP phases in the
$J_1$-$J_2$ model include a columnar dimer VBC \cite{Read1989}, a
plaquette VBC \cite{Zhitomirsky1996}, but also a SL \cite{Capriotti2001}. 
For the N\'eel to VBC transition deconfined quantum criticality has been proposed 
as a novel scenario \cite{Senthil2004,rachid08}. Experimentally, the $J_1$-$J_2$
model may be realized in several layered materials such as
Li$_2$VO(Si,Ge)O$_4$ \cite{LiVO}, VOMoO$_4$ \cite{VOMo}, and
BaCdVO(PO$_4$)$_2$ \cite{BaCdVO}.

\begin{figure}[tb]
\begin{center}
\scalebox{.35}{
\includegraphics{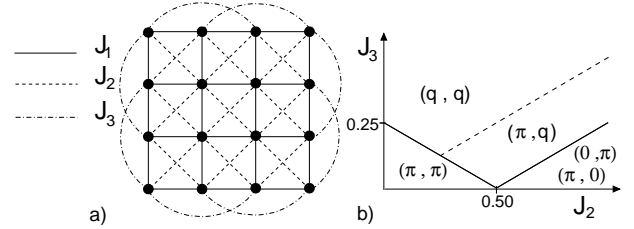}}
\caption{\label{figure1} a) $J_{1}$-$J_{2}$-$J_{3}$ model.
Solid dots refer to lattice sites. Only representative
third-nearest-neighbor exchange paths are depicted. b) The classical
phase diagram of the $J_{1}$-$J_{2}$-$J_{3}$ model.}
\label{fig1}
\end{center}
\end{figure}

One approach to shed additional light on the QP region of the $J_1$-$J_2$ model
is to embed its analysis into a larger parameter space. In this context the \jjj
model
\begin{equation}
\label{ham}
H =
J_1 \sum_{\langle i, j \rangle}{\bf S}_i \cdot {\bf S}_{j}
+J_2 \sum_{\langle\langle i, j \rangle\rangle} {\bf S}_i \cdot {\bf S}_j
+J_3 \sum_{\langle\langle\langle i, j \rangle\rangle\rangle}
{\bf S}_i \cdot {\bf S}_j
\end{equation}
has become of renewed interest recently. ${\bf
S}_i$ refers to spin-$1/2$ operators on the sites of the planar
square lattice shown in Fig.~\ref{figure1} a), and $J_{1,2,3}$ are exchange
couplings ranging from first, i.e. $\langle i, j \rangle$, up to third-nearest
neighbors, i.e. $\langle\langle\langle i, j \rangle\rangle\rangle$. For the
remainder of this work we will focus on the AFM case,
i.e. $J_{1,2,3}\geq 0$ and set $J_{1}=1$.

Classically, the \jjj model allows for four ordered phases
\cite{Moreo1990,Chubukov1991,Rastelli1992,Ferrer1993,Ceccatto1993},
comprising a N\'eel, a collinear, and two types of spiral states which are
depicted in Fig.~\ref{figure1} b). Except for the transition from the diagonal
$(q,q)$-spiral to the $(\pi,q)$-spiral state, which is first order, all
remaining transitions are continuous. Early analysis of quantum fluctuations
\cite{Ferrer1993} found the N\'eel phase to be stabilized by $J_3>0$, with
the end-point of the classical critical line $J_3 = 1/4 - J_2/2$ at $J_2=0$
shifted to substantially larger values of $J_3$. First indications of
non-classical behavior for finite $J_3 >0$ where obtained at $J_2 = 0$. A
'spin-Peierls state' was found in exact diagonalization (ED) studies in the
vicinity of $J_3\sim0.7$, between the N\'eel phase and the diagonal spiral
\cite{Leung1996}. Monte-Carlo and $1/N$ expansion resulted in a succession
of a VBC and a $Z_2$ spin-liquid in this region \cite{Capriotti2004}. QP
behavior was also conjectured at finite $J_2, J_3$, along the line $J_2 = 2J_3$
using Schwinger-Bosons \cite{Ceccatto1993}. More recent analysis, based on
ED and short-range valence bond methods found an s-wave plaquette VBC, breaking
only translational symmetry, along the line $J_2+J_3 = 1/2$, up to $J_2 \lesssim
0.25$ \cite{Mambrini2006}. This VBC's region of stability was then studied
by series expansion in the $(J_2,J_3)$ plane \cite{Arlego2008}.  Results
from projected entangled pair states (PEPS) at $J_2=0$ supported the notion of
an s-wave plaquette along the $J_3$-axis \cite{Murg2009}. However, the
symmetry of the QP state remains under scrutiny, since a truncated quantum-dimer
model \cite{Ralko2009} indicates that the potential plaquette VBC has a
subleading columnar dimer admixture in the vicinity of $J_2\approx J_3\approx
0.25$, similar to ED studies \cite{Sindzingre2010}. This implies broken
translation and rotation symmetry. For $J_2\gtrsim 0.5$, ED shows strong
columnar dimer correlations \cite{Sindzingre2010}. Finally, the order of
the transitions from the QP into the semiclassical phases, and in particular to
the diagonal spiral, remain an open issue.

In this work we intend to further clarify the {\em extent} of the QP regime,
using three complementary techniques, namely, functional renormalization group
(FRG), coupled cluster methods (CCM), and series expansion (SE). These methods
display rather distinct strengths and limitations which we will combine.
CCM and SE are methods which operate inherently in the thermodynamic limit,
however require extrapolation with respect to cluster size or expansion order.
The FRG method is in principle also formulated in the thermodynamic limit, but
its numerical implementation requires one to restrict the spin correlation
length to a maximal value, which is much larger than system sizes in ED
studies. At present neither of these methods alone allows to investigate the
full range of semiclassically ordered and QP states, however their combinations
provides completive information on the quantum critical lines bounding QP
regions: FRG can signal magnetic instabilities of a paramagnetic state, the SE
limits the QP region, and CCM clarifies the stability of part of the ordered
states. As a main result of this paper we will show that the quantum critical
lines agree remarkably well between all three methods, establishing part of
the quantum paramagnetic region rather firmly. Unfortunately none of our
approaches allow to determine the symmetry of the QP state unbiased, which leaves
this an open issue. The paper is organized as follows. In section II we provide
for a brief technical account of all three approaches. Section III is devoted to
a discussion of the results.  We conclude in section IV.

\section{Methods}

In this work we mainly employ three methods to deal with quantum spin systems,
namely FRG, see section~\ref{frg}, which uses a diagrammatic, dynamical
renormalization group approach, CCM, see section~\ref{ccm}, which is a cluster
expansion method employing an exponential ansatz for the correlated ground
state, and finally SE in the exchange coupling constants, see section~\ref{se},
based on continuous unitary transformations. In the following we briefly explain
each of these methods.

\subsection{Functional Renormalization Group Method}
\label{frg}

\begin{figure}[tb]
\begin{center}
\includegraphics[scale=0.5]{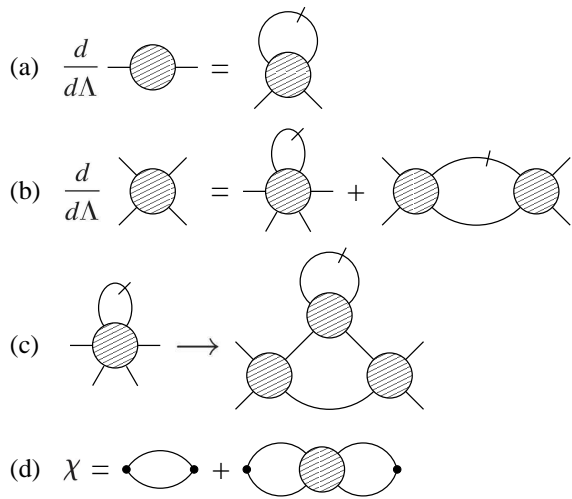}
\caption{The FRG scheme in diagrammatic form. Lines without a slash represent
the Green's functions and lines with a slash the single-scale propagators. The
different vertices are given by circles. The equations (a) and (b) show the FRG
flow equations for the selfenergy and the two-particle vertex,
respectively. Note that Eq. (b) does not distinguish between the
particle-particle channel and the different particle-hole channels. For a full
presentation see e.g. Ref. \onlinecite{reuther10}. The Katanin truncation scheme
is given by the replacement (c). In (d) the relation between the
spin-correlation function $\chi$ and the two-particle vertex is displayed.}
\label{fig2}
\end{center}
\end{figure}

The first approach to tackle the system is based on the functional
renormalization group (FRG) in conjunction with a pseudo-fermion representation
of the $S=1/2$ spin operators. A detailed description of the FRG in general is
given e.g. in Ref.~\onlinecite{salmhofer01}. For an implementation of FRG with
pseudo fermions and an application to the $J_{1}$-$J_{2}$-Heisenberg-model, we
refer the reader to Ref.~\onlinecite{reuther10}. This approach is guided by the
idea to treat spin models in the framewok of standard Feynman many body
techniques. In order to be able to apply the methods of quantum field theory
(Wick's theorem), we use the pseudo-fermion representation of spin operators,
\begin{equation}
S^{\mu}=\frac{1}{2}\sum_{\alpha\beta}f_{\alpha}^{\dagger}
\sigma_{\alpha\beta}^{\mu}f_{\beta}\quad,\quad\alpha,
\beta=\uparrow,\downarrow\quad,\quad\mu=x,y,z\;,
\end{equation}
where $f_{\uparrow}$ and $f_{\downarrow}$ are the annihilation operators of the
pseudo fermions and $\sigma^{\mu}$ are the Pauli-matrices. This representation
requires a projection of the larger pseudo-fermion Hilbert space (4 states per
lattice site) onto the physical subspace of singly occupied states (2
states). At zero temperature we may perform this projection by putting the
chemical potential of the pseudo fermions to zero. Empty or doubly occupied
states are acting like a vacancy in the spin lattice and are therefore
associated with an excitation energy of order $J$. Quantum spin models are
inherently strong coupling models, requiring infinite resummations of
perturbation theory. The simplest such approach is mean-field theory of the spin
susceptibility, which is known to provide qualitatively correct results in the
case that a single type of order is present. On the other hand, frustrated
systems are characterized by competing types of order. This is a situation when
FRG is a powerful tool, as it allows to resum the contributions in all the
different (mixed) channels in a controlled and unbiased way. The first step is
the introduction of a sharp infrared frequency cutoff for the Matsubara Green's
functions. FRG then generates a formally exact hierarchy of coupled differential
equations for the one-particle-irreducible vertex functions where the frequency
cutoff $\Lambda$ is the flow parameter. In Fig.~\ref{fig2} we show the two first
equations of the hierarchy, the first one, Fig.~\ref{fig2}~(a) for the
pseudo-fermion selfenergy, which has a crucial role in particular for highly
frustrated interactions (see Ref.~\onlinecite{reuther10}), the second one,
Fig.~\ref{fig2}~(b), for the two-particle vertex function. The $\beta$-function
of the latter has a contribution given by the three-particle vertex
function. Following Katanin (Ref.~\onlinecite{katanin,salmhofer}) we approximate
the three-particle vertex by the diagram shown in Fig.~\ref{fig2}~(c). In this
way the random phase approximation (RPA) is recovered as a diagram subset, ensuring the
qualitatively correct behavior on the approach to an ordered phase. Another way
of saying this is that the conserving properties of the approximation (the Ward
identities) are satisfied in a better way. It is worth noting that without the
three-particle vertex contribution, RPA cannot be recovered. Keeping the
contribution Fig.~\ref{fig2}~(c) is formally equivalent to the replacement of
the single scale propagator $S^{\Lambda}(i\omega)$ by
$-\frac{d}{d\Lambda}G^{\Lambda}(i\omega)$ (where $G^{\Lambda}(i\omega)$ is the
scale-dependent Green's functions). The approximation may be regarded as a
natural extension of the usual one-loop truncation, in which all
three-particle vertex contributions are discarded. On the other hand it is also
important to keep all the terms consisting of two two-particle vertices on the
r.h.s. of Fig.~\ref{fig2}~(b), as they control disorder tendencies and
therefore the size of the paramagnetic region.

The FRG equations depicted in Fig.~\ref{fig2} (a)-(c) are solved on the
imaginary frequency axis and in real space, rather than in momentum space. The
numerical solution requires a discretization on the frequency axis by a
logarithmic mesh. We found that it is essential to keep the full frequency
dependence of the vertex function (3 frequency variables). The spatial
dependence is approximated by keeping correlation functions up to a maximal
length. As a result, for each set of discrete frequencies and site indices on
external legs of a vertex function, one RG equation is obtained. For well
converged results we typically need to keep sets of about $10^6$ coupled
ordinary differential equations. In the present formulation long-range order
(LRO) is not taken into account. Therefore we should not find a stable solution
of the equations down to $\Lambda=0$ in the parameter regimes where LRO is
present. The existence of a stable solution therefore indicates the absence of
LRO. It is worth to emphasize again that our FRG approach has no bias
concerning magnetic LRO or a paramagnetic state. Our starting point of free,
dispersionless auxiliary fermions does not imply any tendency towards a certain
state.

The physical quantities of interest here, the spin susceptibility and spin
correlation function may be obtained from the diagrams depicted in
Fig.~\ref{fig2} (d). Below we discuss results for the static
susceptibilities as a function of the wave vector. In the ordered phases the
susceptibility at the $k$-vector corresponding to the magnetic LRO is found to
increase as the running cutoff $\Lambda$ is decreased, until the solution
becomes unstable below a certain value of $\Lambda$. Thus, the $k$-vector
characterizing the magnetic order at hand may be determined as that
corresponding to maximal growth of the susceptibility. If the susceptibilities
flow smoothly towards $\Lambda=0$ for any $k$-vector, we are in a disordered
phase.

\subsection{Coupled Cluster Method}
\label{ccm}

Next we analyze the system from a complementary viewpoint using the CCM, the
main features of which we briefly illustrate now. For more details the reader is
referred to Refs.~\onlinecite{zeng98,krueger01,bishop00,farnell04,rachid05} and
references therein.  We mention, that the CCM has been applied successfully to
determine the stability range of magnetically ordered ground state phases in
frustrated quantum magnets
\cite{krueger01,rachid05,Schm:2006,bishop08,bishop08a,zinke08,rachid08,richter10}.
Moreover, it has been demonstrated that the CCM is appropriate to investigate
frustrated quantum spin systems with incommensurate magnetic structures
\cite{bursill,krueger01,rachid05,bishop09,zinke09}.  The starting point for a
CCM calculation is the choice of a normalized reference state $|\Phi\rangle$,
together with a set of mutually commuting multispin creation and destruction
operators $C_I^+$ and $C_I^-$, which are defined over a complete set of
many-body configurations $I$.  We choose $\{|\Phi\rangle;C_I^+\}$ in such a way
that we have $\langle\Phi|C_I^+=0=C_I^-|\Phi\rangle$, $\forall I\neq 0$.  Note
that the CCM formalism corresponds to the thermodynamic limit
$N\rightarrow\infty$.  Depending on the model parameters $J_1$, $J_2$ and $J_3$
we have considered the N\'eel, the collinear, and the diagonal spiral
state. Results on the $(\pi,q)$-state could not be obtained at sufficient
precision. We work in a locally rotated frame of reference such that all spins
of the reference state align along the negative $z$ axis.  Obviously, the choice
of the rotated coordinate frame depends on the choice of the reference state
$|\Phi\rangle$. For a spiral reference state the local rotation angle is related
to the pitch $q$.  In the rotated coordinate frame the reference state reads
$|\Phi\rangle \hspace{-3pt} = \hspace{-3pt} |\hspace{-3pt}\downarrow\rangle
|\hspace{-3pt}\downarrow\rangle |\hspace{-3pt}\downarrow\rangle \ldots \,$, and
we can treat each site equivalently.  The corresponding multispin creation
operators then can be written as $C_I^+=s_i^+,\,\,s_i^+s_{j}^+,\,\,
s_i^+s_{j}^+s_{k}^+,\cdots$, where the indices $i,j,k,\dots$ denote arbitrary
lattice sites.

The CCM is based on ket and a bra ground states, $|\Psi\rangle$ and
$\langle\tilde{\Psi}|$ respectively, which are parameterized as
\begin{eqnarray}
\label{ket1}
|\Psi\rangle = e^S|\Phi\rangle \ , 
\qquad 
S = \sum_{I \neq 0}{\cal S}_IC_I^+ \ , 
\nonumber\\
\langle\tilde{\Psi}| =  \langle\Phi|\tilde{S}e^{-S} \ ,
\qquad 
\tilde{S} = 1 + \sum_{I \neq 0} \tilde{\cal S}_IC_I^- \ ,
\end{eqnarray}
where the so-called correlation coefficients ${\cal S}_I$ and $\tilde{\cal S}_I$
are determined from the CCM equations
\begin{eqnarray}
\label{ket_eq}
\langle\Phi|C_I^-e^{-S}He^S|\Phi\rangle = 0,
\\
\label{bra_eq}
\langle\Phi|\tilde{\cal S}e^{-S}[H, C_I^+]e^S|\Phi\rangle = 0 \ ,
\end{eqnarray}
for each $I$. Using the Schr\"odinger equation, $H |\Psi\rangle = E_0
|\Psi\rangle$, the ground state energy can be written as $E_0 = \langle
\Phi|e^{-S}He^S |\Phi\rangle$, whereas the magnetic order parameter is given by
$m = -\sum_{i=1}^N \langle\tilde\Psi|s_i^z|\Psi\rangle/(Ns)$ , where $s_i^z$ is
expressed in the rotated coordinate frame and $s=1/2$ is the spin quantum
number. We note, that for the spiral state the pitch vector $q$ is used as a
free parameter in the CCM calculation, which has to be determined by
minimization of the CCM ground state energy with respect to $q$.

In order to proceed, the operators $S$ and $\tilde{S}$ have to be truncated
approximately. Here we use the well elaborated LSUB$n$ scheme, where only $n$ or
fewer correlated spins in all configurations, which span a range of no more than
$n$ adjacent (contiguous) lattice sites, are included.  The number of
fundamental configurations can be reduced exploiting lattice symmetry and
conservation laws. In the CCM-LSUB10 approximation we have finally $29\,605$
($45\,825$) fundamental configurations for the N\'eel (collinear) reference
state and for the CCM-LSUB8 approximation we have finally $20\, 876$ fundamental
configurations for the spiral reference state.

To obtain results at $n\rightarrow\infty$, the 'raw' LSUB$n$ data have to be
extrapolated. While there are no a-priori rules to do so, a great deal of
experience has been gathered for the GS energy and the magnetic order parameter.
For the GS energy per spin $E_0(n) = a_0 + a_1(1/n)^2 + a_2(1/n)^4$ is a
reasonable well-tested extrapolation ansatz
\cite{rachid08,krueger01,Schm:2006,rachid05,farnell04,bishop00,bishop08,bishop08a}.
An appropriate extrapolation rule for the magnetic order parameter for systems
showing a GS order-disorder transition is
$m(n)=b_0+b_1(1/n)^{1/2}+b_2(1/n)^{3/2}$ with {\em fixed} exponents, see Refs.
\onlinecite{bishop08,bishop08a,zinke08,richter10}. Extrapolations
$m(n)=c_0+c_1(1/n)^{c_2}$, with a {\em variable} exponent $c_2$ have also been
employed \cite{rachid08,Schm:2006,rachid05,zinke08}.

\subsection{Series Expansion}\label{se}

Finally we highlight SE as the third approach which we employ.  
Our SE for the \jjj model will not be carried out on Eq.~(\ref{ham}), but on a
Hamiltonian which is obtained by a continuous unitary transformation (CUT)
\cite{Wegner1994,Knetter2000}. This transformation is designed such as to
pre-diagonalize the Hamiltonian with respect to a discrete {\em 'particle'
number} $Q$ which counts the number of excitation quanta within an eigenstate of
the {\em unperturbed} spectrum. Therefore the SE can be carried out in spaces of
fixed $Q$, which greatly reduces the computational complexity as compared to
other SE methods. For the latter notions to be of reason, the unperturbed energy
spectrum has to be equidistant, which limits the particular types of unperturbed
Hamiltonians and phases which can be analyzed. Here we will consider CUT SE
results for the \jjj model which have been obtained from an unperturbed
Hamiltonian which leads to a {\em plaquette} VBC ground state. I.e., in contrast
to the CCM, the SE starts from the disordered phase. Using a plaquette
VBC for this phase is motivated by results from ED, short-range resonating
valence bond methods \cite{Mambrini2006,Sindzingre2010}, and truncated
dimer models \cite{Ralko2009} which suggest that plaquette correlations in
the QP state are dominant for $J_2<0.5$.  The plaquette VBC will break only
translational symmetry. Additional, subleading columnar correlations, which have
also been found recently \cite{Ralko2009,Sindzingre2010} are not
included. Some SE results for the phase diagram of the \jjj model have been
given in Ref.~\onlinecite{Arlego2008}. Here we extend this analysis by also
calculating the ground state energy and by comparison with FRG and CCM. CUT SE
using plaquette VBCs has been carried out for various systems recently
\cite{Arlego2008,Arlego2007,Arlego2006,Brenig2004,Brenig2003,Brenig2002}
and we refer the reader to there, for more details. To start, the Hamiltonian is
decomposed into
\begin{equation}\label{mw1}
H = H_0 + H_1 = \sum_{\bf l} 
H^\Box_{\bf l}(J_2) + 
\sum_{{\bf l},{\bf m}} H^\Box_{{\bf l},{\bf m}}(J_1,J_2,J_3),
\end{equation}
where the first sum refers to a dense partitioning of the lattice in
Fig.~\ref{fig1} a) into disjoint four-spin plaquettes, diagonally crossed by two
$J_2$ couplings and with $J_1$ on the plaquettes set to unity. $H_0(J_2=0)$ has
an equidistant energy spectrum.  The second sum contains all inter-plaquette
couplings, with $(J_1,J_2,J_3)$ being the expansion parameters of the SE. After
the CUT the Hamiltonian reads
\begin{equation}\label{Heff}
    H_{\mathrm{eff}} = H_0 + \sum _{k, m, l=1}^\infty  C_{k,m,l} J_1^k
J_2^m J_3^l\,\,,
\end{equation}
where each $C_{k,m,l}$ are sums of weighted products of local and
inter-plaquette operators $O_i^n$, which create ($n \geq 0$) and destroy ($n <
0$) quanta due to $J_i$ within the ladder spectrum of $H_0(J_2=0)$. The weights
are fixed by $H_0$ from a set of differential flow-equations \cite{Knetter2000}
and the $O_i^n$ are evaluated once for a given topology of exchange couplings
$J_{1,2,3}$. We note that for the \jjj model we find $|n|\leq 4$. The main point
is that the {\em total} number of quanta generated by each addend in the sum in
Eq.~(\ref{Heff}) is {\em zero}. In turn, the eigenstates are classified by $Q$
and their energy is obtained by diagonalizing $H_{\mathrm{eff}}$ within an $N^Q$
dimensional space only, where $N$ is the system size. For the ground state
energy $Q=0$ and $N^Q\equiv d=1$, which implies that it is given by a single matrix
element of $H_{\mathrm{eff}}$, namely $E_0=\langle 0|H_{\mathrm{eff}}| 0\rangle$, where
$| 0\rangle$ is the {\em unperturbed} bare plaquette state. For one-particle
excitations $Q=1$ and $d=N$, which however, due to translational invariance also
reduces to $d_{\bf k}=1$, where ${\bf k}$ refers to momentum. Both, the weights
and the operators in (\ref{Heff}) can be evaluated exactly by arbitrary
precision arithmetic codes, leading to {\em analytic} results.

\section{Results}

\subsection{Ground State Energy}

\begin{figure}[tb]
\includegraphics[width=0.89\columnwidth]{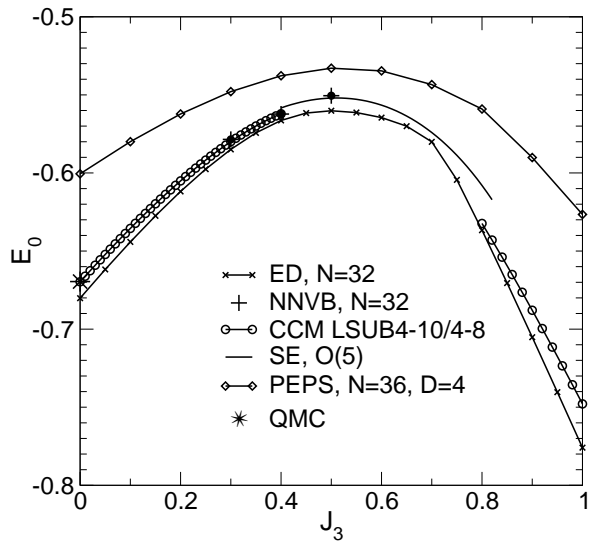}
\caption{Ground state energy at $J_2=0$:  CCM, SE, and ED from this work,
nearest-neighbor valence-bond (NNVB) basis from Ref.~\onlinecite{Mambrini2006},
PEPS from Ref.~\onlinecite{Murg2009}, and quantum Monte-Carlo (QMC) from
Ref.~\onlinecite{Kim1998}. CCM extrapolation see text. All energies are given 
in units of $J_1$.}
\label{fig3}
\end{figure}

We have used CCM and SE to calculate the ground state energy $E_0$ in the
ordered and the QP phase, respectively. The ground state energy has also been
obtained from ED on 32 sites, both, within the complete Hilbert space and for a
nearest neighbor valence-bond basis \cite{Mambrini2006}. PEPS calculations have
also reported $E_0$, however for $J_2=0$ only \cite{Murg2009}. It is therefore
instructive to compare results from these various methods. This is shown in
Fig.~\ref{fig3}, where we have extended the ED of Ref.~\onlinecite{Mambrini2006}
by calculating more data points and considering also the ordered regimes at
$J_2=0$. The CCM data in this figure refers to extrapolations using LSUB4-10
(LSUB4-8) in the N\'eel (spiral) state, as detailed in section~\ref{ccm}.  The SE
is calculated to $O(5)$ in $J_{2,3}$ and the inter-plaquette $J_1$. The complete
corresponding analytic expression is to lengthy to be displayed explicitly
\cite{mailSE}, however at $J_1$=1 and $J_2$=0, as in the figure, it reduces to
\begin{eqnarray}
E_0 & = & -\frac{720160379}{1045094400}
+\frac{50524391297 J_3}{87787929600}
-\frac{25907023 J_3^2}{34292160}
\nonumber \\
&& +\frac{2377469 J_3^3}{5598720}
-\frac{198439 J_3^4}{1128960}
-\frac{1704733 J_3^5}{13547520}
\label{egSE}
\end{eqnarray}
The SE result depicted refers to this bare series with no additional
extrapolations performed.

First we note that the classical energy, which is not contained in this figure,
varies similar with $J_3$ as compared to the other graphs plotted, however it is
larger by $\sim 0.17J_1$ on the average. Secondly, it is obvious that PEPS is
significantly higher than the other results \cite{Dscaling}. This discrepancy is
most pronounced in the ordered regimes, where $E_0 =-0.66953(4)$ is a best
currently available value from Quantum Monte-Carlo \cite{Kim1998} for the
nearest neighbor Heisenberg model.  On the other hand, both, CCM and SE are
rather close to the ED. Each of them has been plotted up to the critical values
$J^{c_1}_3,\ J^{c_2}_3$ which define the extend of the ordered and QP phases as
determined in the following section. At the endpoints CCM and SE match up
acceptably, where the agreement at $J^{c_1}_3$ is best and the convergence of
the SE may be less reliable for $J^{c_2}_3$ which is the larger. Thirdly, having
in mind that the finite-size shift of the ED data for $N=32$ is about $+0.01$,
see Refs. \onlinecite{Richter2010,richter10}, it is remarkable that the CCM, ED
and SE data almost coincide if $J_3$ is not too large.  The increase of the
difference between the ED and CCM data at larger $J_3$ can be attributed to the
crossover (i) of the characteristic length scale from nearest-neighbor
($J_1$-bonds) to 3rd-nearest-neighbor ($J_3$-bonds) separation and (ii) of the
characteristic energy scale from $J_1$ to $J_3$.  While the first crossover
effect leads to an enhanced finite-size effect in the ED energy and to a larger
impact of LSUB$n$ clusters with $n$ beyond those considered here, the second
crossover effect automatically enhances any discrepancy of energies roughly
proportional to $J_3$. Finally we note that, also energies obtained from ED
using a restricted nearest neighbor valence-bond basis \cite{Mambrini2006} agree
very well with those from our CCM, SE, and complete Hilbert space ED.

\subsection{Quantum Phase Diagram}

\begin{figure}[tb] \begin{center}
\includegraphics[width=0.89\columnwidth]{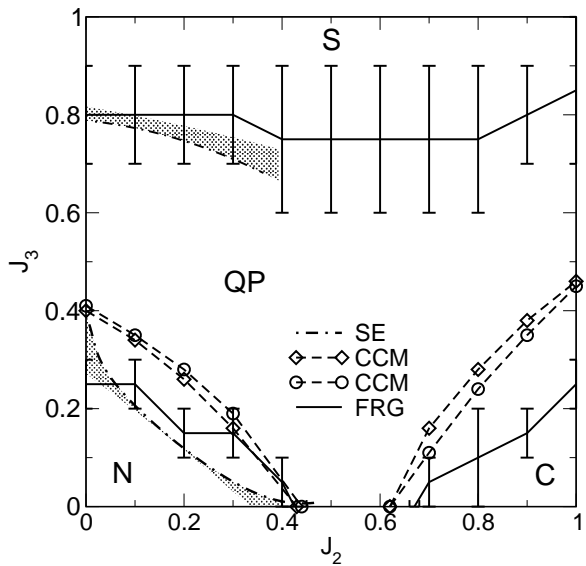}
\caption{Combined quantum phase diagram of the \jjj model.  Solid: Onset of
magnetic flow from FRG on 11$\times$11 sites with 46 frequency points. Error
bars of size 0.1 (larger than 0.1) are due to the finite $J_{2,3}$-mesh
(uncertainties in the flow of the susceptibility). See text.  Dashed: Lines of
vanishing order parameter from LSUB$n$ CCM with fixed(variable)-exponent
extrapolations from n=4,6,8,10 for diamonds(circles). See text.  Dashed-dotted:
Triplet-gap closure from 5th-order CUT SE. Small(Large)-$J_3$ lines are
[3,1]([2,2])-DlogPad\'e approximants. Shaded region refers to difference between
bare series and DlogPad\'es.  'N', 'C', and 'S' denote N\'eel, collinear, and
spiral state. 'QP' refers to a generic quantum paramagnet for CCM and FRG and to
a plaquette VBC for SE.}
\label{fig4}
\end{center}
\end{figure}

Using FRG the phase diagram has been calculated in the $J_2$-$J_3$-plane with
parameter steps of 0.1 for $0\leq J_{2,3}\leq 1$. A large computational effort
is required to solve the system of FRG equations. In the present calculation we
used 46 frequency points. The spatial dependence of the susceptibility was kept
up to lattice vectors $\mathbf{R}$ satisfying $\textnormal{Max}(|R_x|,|R_y|)\leq
5$, and the susceptibilities were put to zero beyond that range. This
provides a correlation area of $11 \times 11$ lattice points, which proved to be
sufficient for a first exploration of the phase diagram. The results were then
Fourier transformed to momentum space. In magnetic phases we see a pronounced
susceptibility peak in momentum space that rapidly grows during the
$\Lambda$-flow. At a certain $\Lambda$ the onset of spontaneous LRO is signalled
by a sudden stop of the smooth flow and the onset of oscillations depending on
the frequency discretization. On the other hand in non-magnetic phases a smooth flow and
broad susceptibility peaks are obtained. This distinction allows us to draw the
FRG phase diagram of the model, which is shown in Fig.~\ref{fig4}. Regarding the
error bars in Fig.~\ref{fig4} we note that bars of size 0.1 into the
$J_3$-direction do {\em not} reflect errors of the FRG, but are only due to
finite $(J_2,J_3)$-spacing and, in principle, apply also to the
$J_2$-direction. However, especially near the phase boundary between the spiral
ordered and the disordered phase, at large $J_3$, we encounter enhanced
uncertainties. Here $(J_2,J_3)$-regions occur where it is not clear if the
behavior of the flow should be interpreted as magnetic or non-magnetic. In
Fig.~\ref{fig4} these regions lead to error bars larger than 0.1.

\begin{figure}[tb]
\begin{center}
\includegraphics[width=0.89\columnwidth]{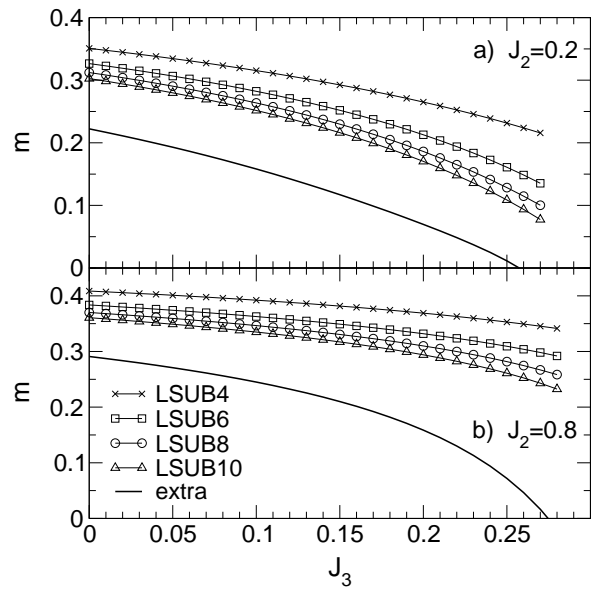}
\caption{Magnetic order parameter $m$ obtained within the CCM-LSUB$n$
approximation as well as extrapolated data using the extrapolation scheme
$m(n)=b_0+b_1(1/n)^{1/2}+b_2(1/n)^{3/2}$ and LSUB$n$ data for $n=4,6,8,10$, see
Sec.\ref{ccm}. a) in the N\'eel phase, b) in the collinear phase.}
\label{fig5}
\end{center}
\end{figure}

To obtain the CCM phase diagram we extrapolate the LSUB$n$ data for the magnetic
order parameter $m$, cf.  section~\ref{ccm}.  Starting in parameter regions
where semiclassical magnetic long range order can be supposed we use the
classical state as the reference state for the CCM. Then we obtain the phase
boundaries of the magnetically ordered phases by determining the lines of
vanishing magnetic order parameter $m$, which implies continuous or second order
transitions. In Fig.~\ref{fig5} we show typical CCM results for order $m$.  For
the N\'eel and the collinear phase we find the extrapolation of $m$ to be nearly
independent of the extrapolation scheme used. Unfortunately, for the spiral
state, computational constraints limit us to LSUB$n$ with $n\leq 8$. Since the
LSUB2 approximation is not appropriate for a proper description at larger $J_3$,
only 3 CCM data points are left for the extrapolation. In that case we find that
the fixed and variable exponent extrapolations lead to rather different critical
values for $J_3$ at fixed $J_2$. Therefore the critical regime of the spiral
state cannot be determined accurately enough from the present CCM. This is very
different for the ground state {\em energy} which allows for stable
extrapolation in all three quasiclassical regions. The location of all
$J^c_3(J_2)$, with $m(J^c_3(J_2)) = 0$, i.e. the quantum phase diagram as
obtained from CCM is included in Fig.~\ref{fig4} for both extrapolation schemes.

Finally the phase boundaries have been calculated using SE. To this end the
plaquette phase has been analyzed with respect to second order instabilities,
i.e., a closure of the elementary triplet gap as a function of $J_2$, $J_3$. For
this, we diagonalize $H_{\mathrm{eff}}$ in the $Q$=1 sector, i.e., the subspace
of single-quanta states $|\mathbf{1}\rangle_{\mathbf{l}}$ at sites ${\bf l}$.
These states are triplets. The sole action of $H_{\mathrm{eff}}$ on these states
is a translation in real space,
$H_{\mathrm{eff}}|\mathbf{1}\rangle_{\mathbf{0}}= \sum_{\mathbf{l}}
c_{\mathbf{l}} |\mathbf{1}\rangle_{\mathbf{l}}$ with the $c_{\mathbf{l}}$
determined from the SE. Fourier transformation yields the triplet dispersion,
similar to a generalized tight-binding problem, with the hopping matrix elements
determined from $H_{\mathrm{eff}}$.  For technical details we refer to
Ref.~\onlinecite{Arlego2008}. We have used this technique to calculate the
triplet dispersion up to 5th order in all three variables $J_1$, $J_2$ and
$J_3$. In Fig.~\ref{fig4} we show the resulting lines for the closure of the
triplet gaps, as obtained from a Dlog-Pad\'e analysis of this dispersion. The
shaded error-region refers to the distance between the critical lines from the
bare SE and those from Dlog-Pad\'e and are a measure of convergence of the
SE. For $J_2\gtrsim 0.5$ the SE's convergence is insufficient to obtain reliable
triplet dispersions.

Figure~\ref{fig4} is a main result of our paper. Most obviously, it shows that
within the range of $J_{2,3}$ investigated, the \jjj model displays a large QP
region. This region extends well beyond the line $J_2+J_3 = 1/2$, with $J_2
\lesssim 0.25$ studied in Ref. \onlinecite{Mambrini2006}, or the vicinity of the
point $J_2\approx J_3 \approx 0.25$ in Ref. \onlinecite{Ralko2009}, and for
$J_2\lesssim 0.5$ also covers a larger $J_3$ interval than that observed in ED
\cite{Sindzingre2010}. The quantum N\'eel phase is enlarged with respect to the
classical one, which agrees with early 1/S-analysis \cite{Ferrer1993} and recent 
ED results\cite{Sindzingre2010}. Our computational approaches are not capable of an
unbiased identification of the symmetry of the QP state. However, since the
phase boundaries predicted from the plaquette SE and those from CCM and FRG
agree rather well, our results corroborate substantial plaquette correlations in
the QP phase for $J_2\lesssim 0.3\ldots 0.4$, which is in line with
Refs. \onlinecite{Mambrini2006,Murg2009,Ralko2009,Sindzingre2010}.

\subsection{Short Range Correlations}

\begin{figure}[tb]
\begin{center}
\includegraphics[width=0.89\columnwidth]{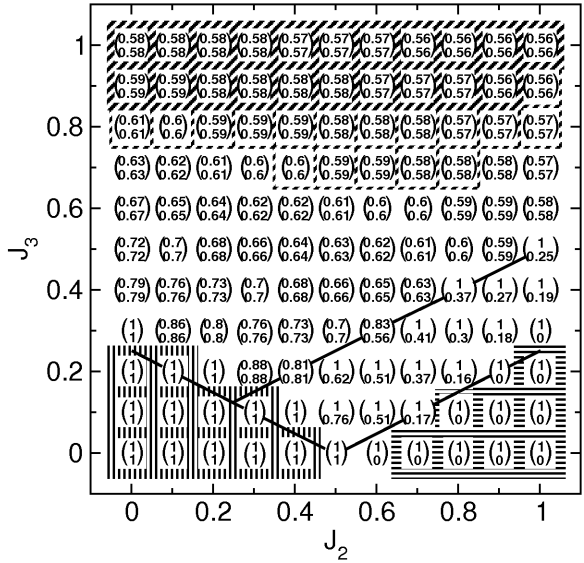}
\caption{Brackets: wave vector $k_x$, $k_y$ in units of $\pi$ at maximum of the
static susceptibility from FRG. Solid lines indicate the classical critical
lines. Vertically, horizontally and diagonally striped frames around the
brackets correspond to N\'eel, collinear, and spiral state, respectively. Thin
frames mark regions of uncertain flow behavior.}
\label{fig6}
\end{center}
\end{figure}

Since FRG evaluates the static susceptibility over the complete Brillouin zone,
it allows to determine the wave vector $\mathbf{k}_{\mathrm{max}}$ of the dominant
short-range magnetic correlations or the pitch vector of the magnetic order
parameter. These wave vectors are depicted in Fig.~\ref{fig6} together with the
quantum phases discussed already in Fig.~\ref{fig4}. Both, in the ordered as
well as in the QP phase we find the wave vectors at maximum of the
susceptibility to agree approximately with those obtained for the purely
classical model in Fig.~\ref{fig1}~b). This is particularly interesting with
respect to the $(\pi,q)$-spiral state, which seems to exist only in the form of
short range correlations in Fig.~\ref{fig6}.

\begin{figure}[tb]
\begin{center}
\includegraphics[width=0.95\columnwidth]{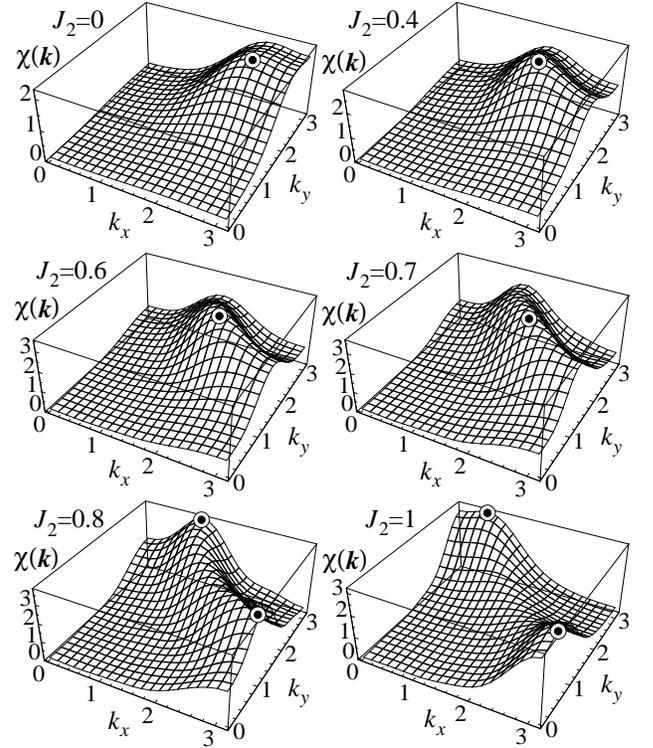}
\caption{Static susceptibility for wave vectors $k_x,k_y\in[0,\pi]$ for various
values $J_2$ and constant $J_3=0.4$. The black dots mark the positions of the
maxima.}
\label{fig7}
\end{center}
\end{figure}

In order to illustrate how the dominant fluctuations in the disordered phase
change with varying couplings, we show in Fig.~\ref{fig7} results for the static
susceptibility as a function of the $k$-vector in the Brillouin zone with
$k_x,k_y\in [0,\pi]$ at fixed $J_3=0.4$ for various values of $J_2$. For $J_2=0$
we see a broadened peak at a $(q,q)$-position which has already moved away from
the N\'eel-point $\mathbf{k}=(\pi,\pi)$. This peak further moves along the
Brillouin-zone diagonal for increasing $J_2$. For $J_2\gtrsim 0.6$ it is seen
that the peak smoothly deforms into an arc and that the weight at the
Brillouin-zone boundary increases. Between $J_2=0.7$ and $J_2=0.8$, close to the
classical first order transition, the ridge has constant weight and the maximum
jumps to a $(\pi,q)$-direction to then further evolve towards the collinear
points $\mathbf{k}=(0,\pi)$, $\mathbf{k}=(\pi,0)$ and to acquire more
prominence. Therefore, remnants of the classical correlations survive into the
QP regime. Very similar behavior is evidenced by ED \cite{Sindzingre2010}.

\section{Conclusion}

To summarize, we have studied the quantum phases of the frustrated planar
$J_1$-$J_2$-$J_3$ spin-1/2 quantum-antiferromagnet, using FRG, CCM, and CUT
SE. This includes evaluations of momentum resolved susceptibilities, the ground
state energy, magnetic order parameters, and the elementary excitation gaps. Our
results provide clear evidence for a sizeable quantum paramagnetic region which
opens up between the N\'eel, collinear, and spiral state of the purely classical
model. A long-range ordered collinear spiral phase, which is also present
classically has not been observed in the quantum model in the parameter range we
have investigated. Where applicable, the agreement between the critical lines
determined from all three methods is remarkably good. While our computational
approaches cannot determine potentially broken symmetries in the quantum
paramagnetic state, the fact that the critical lines which we have obtained from
FRG and CCM agree well with those from the CUT SE which is based on a plaquette
VBC, is indicative of VBC ordering with substantial plaquette correlations in
the quantum paramagnetic region - in those parameter ranges where CUT SE
applies. Our results are consistent with second order transitions from the
N\'eel and the collinear state into the quantum paramagnet. Unexpectedly, our
CCM results do not provide a definite signal of a transition from the spiral
state into the quantum paramagnet. This may simply be related to an insufficient
order of the LSUBn approximation, but could also hint at a first order
transition and an intermediate phase between the VBC and the spiral
state. Finally, while our findings do not show a collinear spiral state as in
the classical model, FRG convincingly demonstrates that the latter is replaced
by corresponding short range correlations in the quantum paramagnetic region.

\section{Acknowledgments}

J.~Reuther and P.~W\"olfle acknowledge support through the DFG research unit FOR
960 "Quantum phase transitions".  M.~Arlego and W.~Brenig acknowledge partial
support by the CONICET through grant 2049/09 and the DFG through grant 444
ARG-113/10/0-1.  J.~Richter and W.~Brenig thank the Max-Planck-Institute for
Physics of Complex Systems for its hospitality during the Advanced Study Group
"Unconventional Magnetism in High Fields" where part of this work was completed.

\end{document}